\documentclass{article}

\usepackage{microtype}
\usepackage{graphicx}
\usepackage{subcaption}
\usepackage{booktabs} % for professional tables
\usepackage{enumitem}
\usepackage{comment}

\usepackage{hyperref}

 \usepackage[preprint]{icml2026}

\usepackage{amsmath}
\usepackage{amssymb}
\usepackage{mathtools}
\usepackage{amsthm}

\usepackage[capitalize,noabbrev]{cleveref}

\theoremstyle{plain}

\theoremstyle{definition}

\theoremstyle{remark}

\usepackage[textsize=tiny]{todonotes}
\usepackage{xspace}

\usepackage{xcolor}

\newcommand{\term}[1]{\texttt{#1}}
\newcommand{\sys}{\term{Kairos}\xspace}
\icmltitlerunning{SLO-Aware Scheduling for Disaggregated LLM Inference}

\begin{document}

\twocolumn[
  \icmltitle{Taming Request Imbalance: \\
  SLO-Aware Scheduling for Disaggregated LLM Inference}



  % \newcommand{\icmlEqualContribution}{\textsuperscript{*}Equal contribution }
  \icmlsetsymbol{equal}{*}

  \begin{icmlauthorlist}
    \icmlauthor{Qipeng Wang}{}
    % \icmlauthor{Firstname3 Lastname3}{comp}
    % \icmlauthor{}{sch}
  \end{icmlauthorlist}

%  \icmlaffiliation{kuaishou}{Kuaishou Inc.}
%  \icmlaffiliation{pku}{Peking University}
%  \icmlaffiliation{ustc}{University of Science and Technology of China}

  \icmlcorrespondingauthor{Qipeng Wang}{861026685@qq.com}

  \icmlkeywords{Machine Learning, ICML}

  \vskip 0.3in
]



\printAffiliationsAndNotice{}  % no special notice (required even if empty)

\begin{abstract}

%Large language model (LLM) serving in production environments must meet stringent service-level objectives (SLOs) under highly variable request patterns. Request lengths in practice follow a long-tail distribution, causing head-of-line blocking on the prefill side and straggler-induced underutilization on the decode side in disaggregated serving architectures. Existing systems, which rely on first-come-first-served prefill scheduling and continuous batching for decode, fail to adapt to this imbalance, leading to degraded SLO attainment and throughput.
%
%We present \sys, an SLO-aware scheduling system that addresses these challenges through two complementary mechanisms. On the prefill side, \sys employs urgency-based priority scheduling, which predicts prefill completion times and dynamically selects requests to maximize time-to-first-token (TTFT) SLO attainment. On the decode side, \sys introduces slack-guided adaptive batching, which exploits the gap between per-step decode time and the time-per-output-token (TPOT) SLO to greedily pack short requests and maximize throughput without violating SLOs. We implement \sys and evaluate it on an online serving dataset and state-of-the-art LLM. Experimental results show that \sys improves TTFT SLO attainment by up to 23.9\%, TPOT SLO attainment by up to 27.1\%, end-to-end SLO attainment by up to 33.8\%, and decode throughput by up to 19.3\% compared to state-of-the-art baselines.

In production environments, large language model (LLM) serving is required to meet stringent service-level objectives (SLOs) amid highly variable request patterns. In practice, request lengths follow a long-tail distribution, which gives rise to head-of-line blocking on the prefill side and underutilization caused by stragglers on the decode side in disaggregated serving architectures. Current systems, which adopt first-come-first-served (FCFS) scheduling for prefill and continuous batching for decode, lack the ability to adapt to this imbalance, resulting in compromised SLO attainment and reduced throughput.

To address these challenges, we propose \sys, an SLO-aware scheduling system equipped with two complementary mechanisms. On the prefill side, \sys employs urgency-based priority scheduling: it predicts prefill completion times and dynamically selects requests to maximize the attainment of time-to-first-token (TTFT) SLOs. On the decode side, \sys introduces slack-guided adaptive batching, which leverages the gap between per-step decode time and the time-per-output-token (TPOT) SLO to greedily pack short requests. This approach maximizes throughput while strictly adhering to SLO requirements. We implement \sys and conduct evaluations using an online serving dataset and a state-of-the-art LLM. Experimental results demonstrate that, compared with state-of-the-art baselines, \sys improves TTFT SLO attainment by up to 23.9\%, TPOT SLO attainment by up to 27.1\%, end-to-end SLO attainment by up to 33.8\%, and decode throughput by up to 19.3\%. 

\end{abstract}
\section{Introduction}

% ---------------- story line----------------

% 1. LLM

% 2. Online serving, SLO and PD disaggregation.

% 3. Problem: request are imbalance, i.e., long-tail distribution, leading to HOL blocking.

% % 3.1. two figures that shows the distributions of the online serving data and a benchmark dataset.

% 4. Existing solution: FCFS + continuous batching. Fail to adapt to such scenario. (1) prefill: long requset blocks the following short request; (2) decode: short request must wait for long request, because of the need for synchronization and different computation complexity of attention at different sequence length.

% 5. Key observation: (1) prefill: prefill time is predictable; (2) decode: SLO is typically larger than the decoding step time and the amount of long request is small.

% 6. Our solution: SLO-aware scheduling

% 6.1 dynamically select the request to prefill based on the urgency, to maximum the TTFT SLO attainment

% 6.2 dynamically batching the request based on profiled decoding step time to maximum the decoding throughput.

% 7. Contributions:

% 7.1 Identify the challenges of  underutilization in online serving system

% 7.2 propose a SLO-aware scheduling method to maximum the SLO attainment and system throughput

% 7.3 Experiments shows that the our system is efficient.

% --------------------------------------------

Large language models (LLMs) have demonstrated remarkable capabilities across a wide range of tasks~\cite{vaswani2017attention, achiam2023gpt, liu2024deepseek, zhao2023survey}, from natural language understanding and generation to code synthesis and mathematical reasoning. Their strong performance has made them a foundational technology for numerous real-world applications.

Serving LLMs in production environments demands highly efficient systems to meet stringent service-level objectives (SLOs)~\cite{wu2024loongserve}. LLM inference consists of two distinct phases: the prefill phase, which processes the input prompt and generates the initial key-value cache, and the decode phase, which iteratively generates output tokens. These phases are governed by two critical SLO metrics: time-to-first-token (TTFT) for prefill and time-per-output-token (TPOT) for decode. To prevent interference between them and maximize system goodput, modern serving systems commonly adopt a prefill-decode disaggregation architecture~\cite{zhong2024distserve}, where the two phases are handled by separate hardware resources.

A fundamental challenge in online LLM serving stems from the highly variable and unpredictable lengths of incoming requests~\cite{agrawal2023sarathi}. With modern LLMs supporting context windows spanning up to hundreds of thousands of tokens~\cite{achiam2023gpt, deepseekv4}, request lengths in practice follow a long-tail distribution: the vast majority of requests are short, while a small fraction are extremely long~\cite{ikram2025ascendra, wu2023fast}. This imbalance leads to severe inefficiencies in disaggregated architectures, affecting both the prefill and decode phases in distinct ways.

On the prefill side, existing disaggregated systems typically adopt first-come-first-served (FCFS) scheduling~\cite{zheng2024sglang, kwon2023efficient}. Under this policy, a long request that arrives ahead of subsequent short requests can monopolize the prefill stage, leading to head-of-line blocking~\cite{ikram2025ascendra}, substantially inflating their TTFT. On the decode side, continuous batching is widely used~\cite{kwon2023efficient}. However, when short and long requests coexist in the same batch, synchronization barriers at each decoding step force short requests to wait for long requests to finish, even though the short ones require far less computation. As per our analysis in \S\ref{sec:request-imbalance}, the per-step decode time grows significantly with request length, because the complexity of attention operation grows linearly according to the sequence length at the decoding stage. When a short request is batched with a much longer one, it must idle through substantially longer per-step latencies, leaving compute resources underutilized and degrading overall throughput.

\textbf{Our design}. To address the above challenges, we propose \sys. \sys is built upon two key observations: (1) prefill time is highly predictable from input length, queue size and arrival time, enabling scheduling decisions to be made based on estimated completion time; and (2) the TPOT SLO is agnostic to request length—the system only needs to deliver each token within the SLO budget—and the SLO is typically significantly larger than the per-step decode time, leaving a slack buffer that can be exploited.

To mitigate head-of-line blocking on the prefill side, where early-arriving long requests delay subsequent short ones, \sys incorporates urgency-based priority scheduling. Specifically, \sys predicts the prefill finish time for all queued requests first, and computes an urgency score for each based on the SLO and arrival time. It then greedily selects requests for prefill in order of urgency, maximizing TTFT SLO attainment.

To address the straggler problem on the decode side, where short requests are forced to wait for long ones within the same batch and decode step, \sys employs an slack-guided adaptive scheduling that maximizes throughput while respecting SLO constraints. Concretely, \sys profiles the per-step decode time and exploits the slack between decode step time and the TPOT SLO budget during serving. Once the slack is large enough, \sys greedily selects short requests whose decode step time is smaller than the slack, packing them together and decoding them within the slack to maximize decode throughput without violating the TPOT SLO.

\textbf{Evaluation}. We implement a prototype of \sys. We evaluate \sys on an online serving dataset and a state-of-the-art LLM, and the experimental results demonstrate its effectiveness. Under the same request rate, \sys improves the TTFT SLO attainment by up to 23.9\%, TPOT SLO attainment by up to 27.1\%, end-to-end SLO attainment by up to 33.8\%, and decode throughput by up to 19.3\%. 

\textbf{Contributions} are summarized as follows:
\begin{itemize}[leftmargin=*, itemsep=-5pt, topsep=-2mm]
    \item We identify the key challenges of request imbalance and its impact on system underutilization in online LLM serving systems with disaggregated architectures.
    \item We propose \sys, an SLO-aware scheduling method that dynamically prioritizes prefill requests based on urgency and adaptively batches decode requests to maximize both SLO attainment and system throughput.
    \item We implement and evaluate \sys on various tasks, demonstrating significant improvements in SLO attainment and decode throughput.
\end{itemize}

\section{Background and Motivation}

\subsection{LLM Serving System}
LLM inference is a two-phase process consisting of the prefill stage and the decode stage. During prefill, the model processes the entire input prompt in a single forward pass and produces the initial token along with the key-value (KV) cache. During decode, the model generates subsequent tokens one by one in an autoregressive manner, each step attending to all previously generated KV cache. These two phases have fundamentally different computational characteristics: prefill is compute-bound, while decode is memory-bound. More importantly, the two phases interfere with each other when colocated on the same hardware—prioritizing either one can starve the other, causing SLO violations. To mitigate this interference, modern LLM serving frameworks such as SGLang~\cite{zheng2024sglang} and vLLM~\cite{kwon2023efficient} adopt a prefill-decode (PD) disaggregation architecture, where the two phases are deployed on separate sets of hardware resources. In this architecture, prefill instances typically employ chunked prefill~\cite{agrawal2023sarathi} to bound the peak memory usage of extremely long sequences, while decode instances leverage continuous batching~\cite{kwon2023efficient} to maximize output token throughput.

\subsection{Request Imbalance}
\label{subsec:request-imbalance}

% \begin{figure}[ht]
%   \vskip 0.2in
%   \begin{center}
%     \centerline{\includegraphics[width=0.48\columnwidth]{fig/request-length-distribution.pdf}}
%     \caption{
%       Reuqest length distribution of 4 datasets.
%     }
%     \label{fig:request-length-distribution}
%   \end{center}
% \end{figure}

\begin{figure}[t]
	\centering					
	\begin{minipage}[b]{0.22\textwidth}
		\includegraphics[width=1\textwidth]{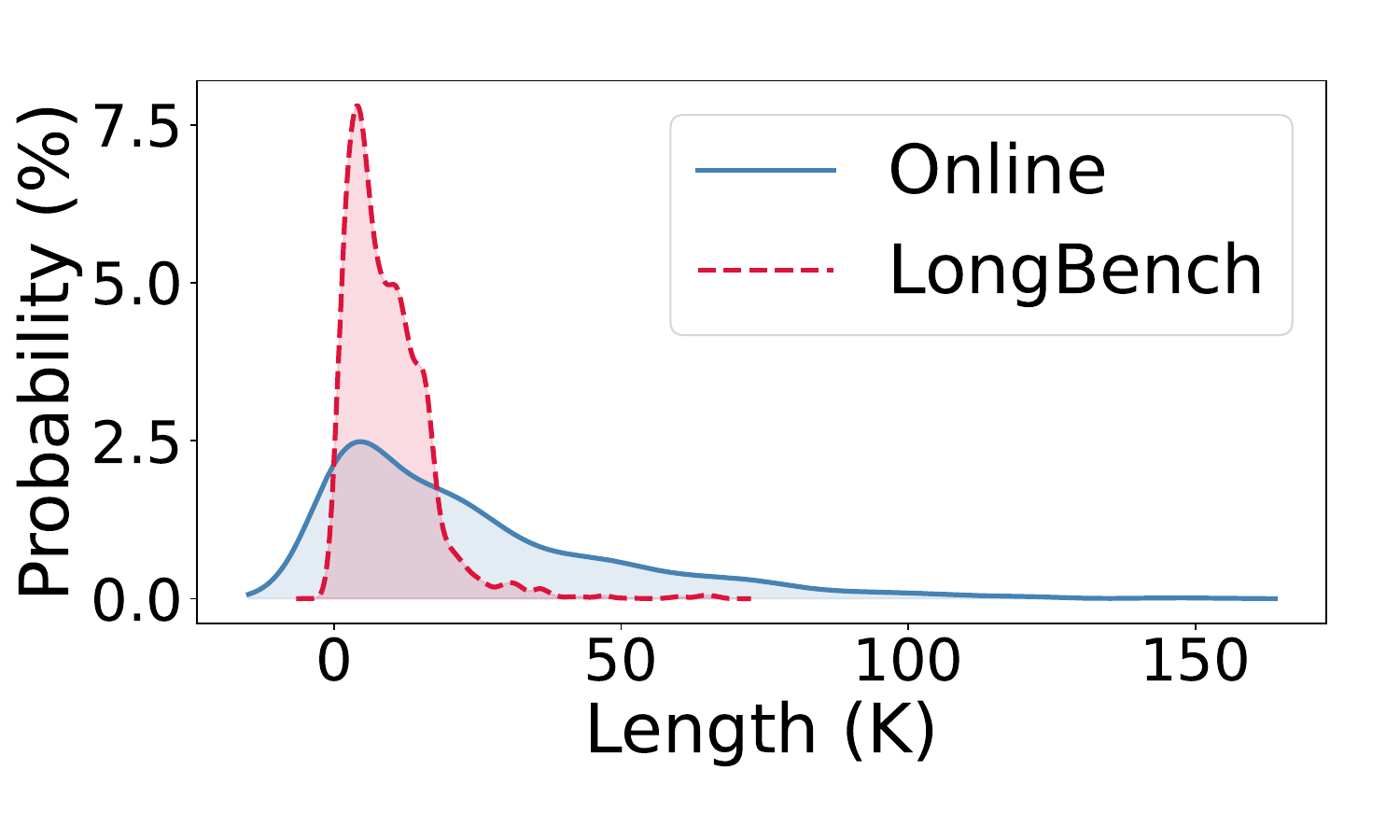}
		\subcaption{Reuqest length distribution of 2 datasets.}
        \label{fig:request-length-distribution}
	\end{minipage}	
	~							
	\begin{minipage}[b]{0.22\textwidth}
		\includegraphics[width=1\textwidth]{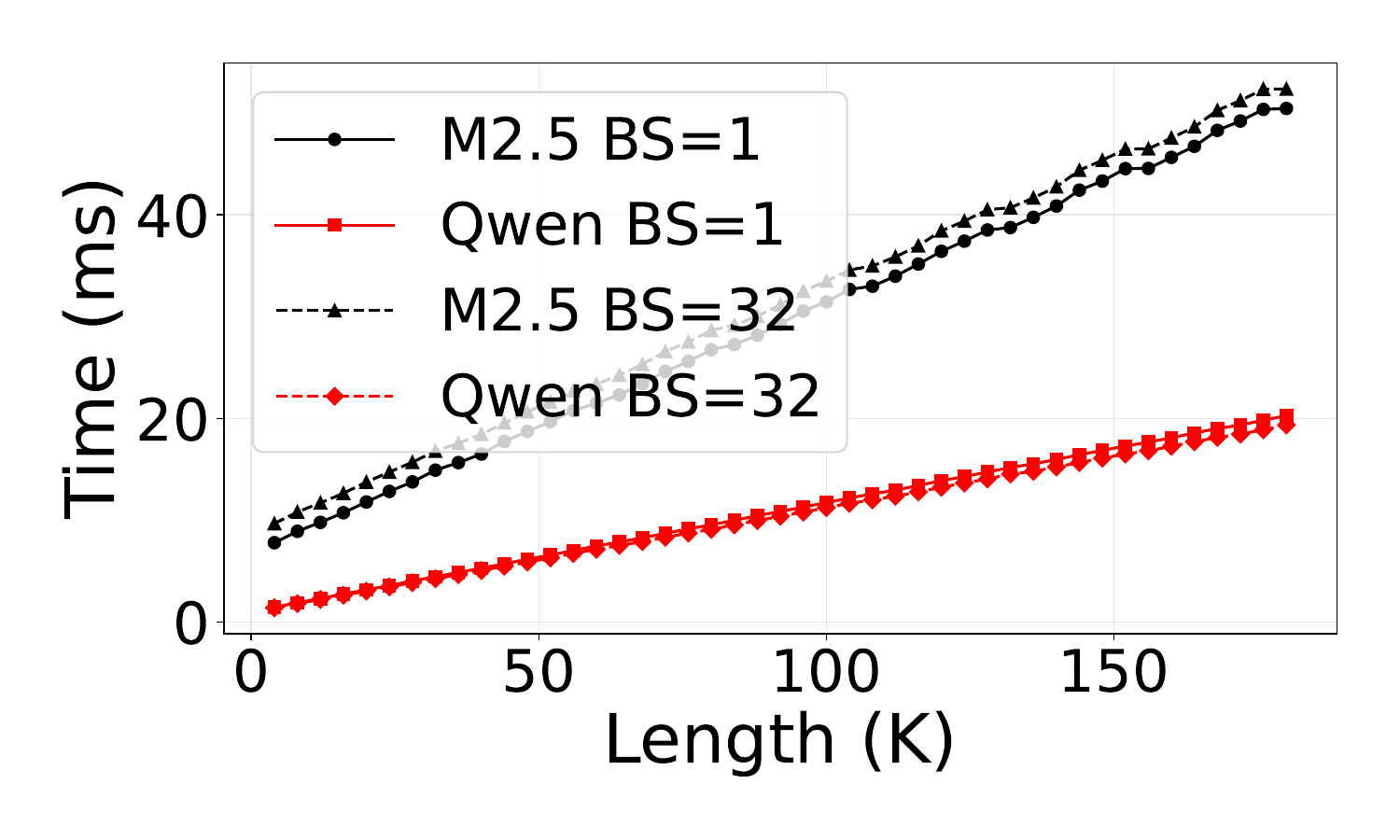}
		\subcaption{Decode step time at different sequence length.}
        \label{fig:decode-time-growth}
	\end{minipage}	
\vspace{-5pt}
\caption{Imbalabce request distribution and decode step time growth to sequence length.}
\label{fig:imbalabce-request-distribution-and-its-impact}
\end{figure}

\begin{figure*}[ht]
	\centering					
	\includegraphics[width=0.95\textwidth]{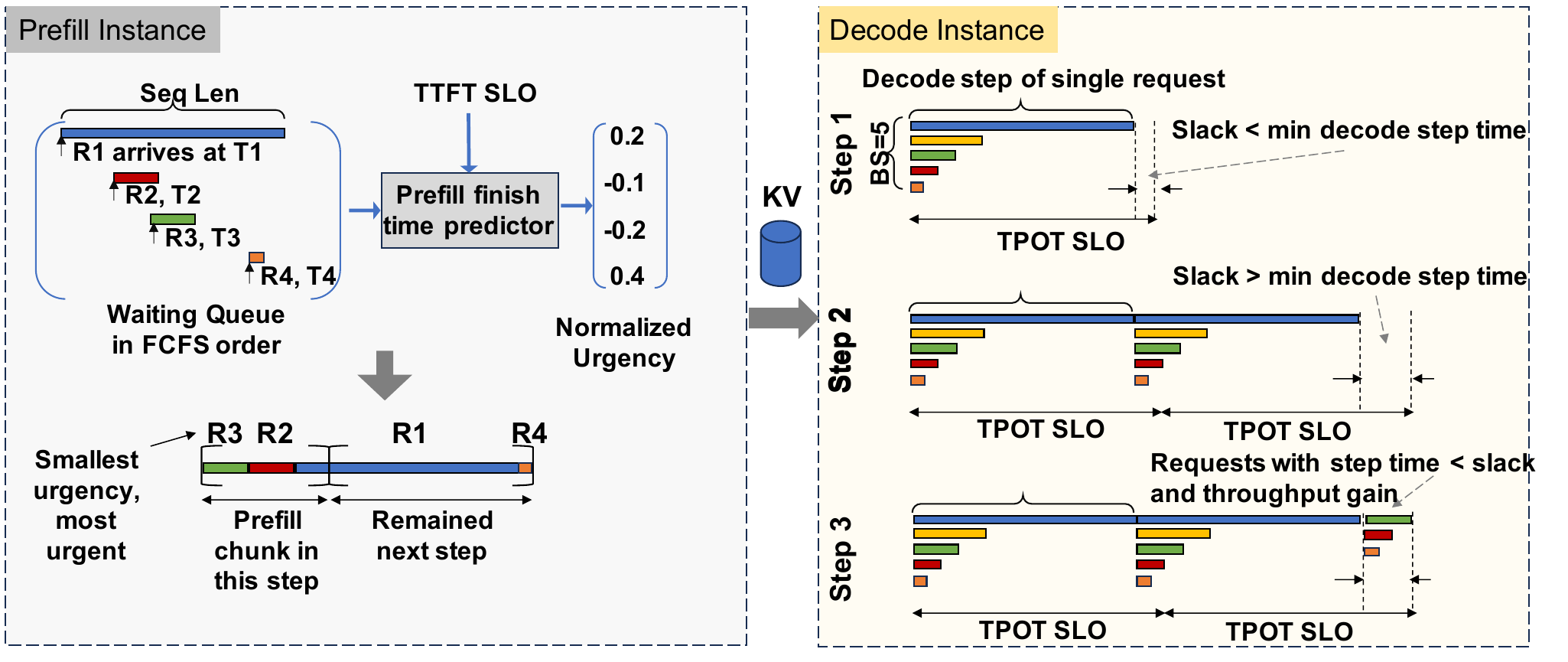}
	\caption{Workflow of the algorithms in \sys.}
	\label{fig:arch}
%	\vspace{-10pt}
\end{figure*}

\textbf{Long-tail distribution of request lengths.} Modern LLMs support increasingly wide context windows, with many models reaching up to 1M tokens~\cite{deepseekv4, claude-opus-4-7}. In practice, the distribution of request lengths in online serving exhibits a pronounced long-tail pattern: the overwhelming majority of requests are relatively short, while a small fraction are extremely long. We observe this phenomenon in both production serving traces and commonly used benchmark datasets~\cite{bai2023longbench}. Figure~\ref{fig:request-length-distribution} illustrates the request length distributions, where the x-axis represents request length and the y-axis shows the cumulative distribution or probability density. The long-tail characteristic is clearly visible in both settings.

\textbf{Impact of request imbalance.} This imbalance leads to severe SLO degradation in both the prefill and decode phases. 
On the prefill side, when a long request arrives ahead of subsequent short requests under FCFS scheduling, it can block others for an extended period, causing their TTFT to exceed the SLO. Under the experimental settings described in \S\ref{subsec:experiment-settings}, a 128K-token request requires approximately 8.8s for prefill, whereas a 8K-token request completes in only  400.4ms for MiniMax-M2.5~\cite{minimax-m2.5}. Consequently, if a 128K-token request arrives first, the maximum sustainable QPS without violating the SLO for the following 8K-tokens requests drops below 0.114. 
On the decode side, the per-step decode time grows linearly with sequence length even at a batch size of one. As shown in Figure~\ref{fig:decode-time-growth}, a 8K-token request takes roughly 11.0ms per decode step, while a 128K-token request incurs approximately 40.3ms. When both coexist in the same batch, the batch step time is determined by the slowest request—namely, 40.3ms. For the 8K-token short request, this means 29.3ms of idle waiting per step due to the synchronization barrier, representing a substantial waste of compute resources.

\subsection{Opportunities} 

Despite the aforementioned challenges, we identify two key opportunities that enable more efficient scheduling under imbalanced workloads.

The first is that the prefill time of a request is highly predictable from its input length and the model architecture~\cite{du2025prefillonly}. This predictability allows the system to estimate, before execution, how long each queued request will take to prefill. When a long request threatens to delay subsequent short requests and degrade TTFT attainment, the scheduler can leverage these predictions to make informed decisions—selecting the request that best balances SLO compliance and overall throughput, rather than blindly following the arrival order.

The second is that there are time slack between TPOT SLO and decode step time. The TPOT SLO is inherently agnostic to request length: the system only needs to deliver one token per SLO interval to meet the target. For instance, with a 50ms TPOT SLO, it suffices to output a token every 50ms. If the system can generate tokens faster than the SLO requires, the excess tokens can be buffered and released gradually, effectively decoupling generation speed from token delivery. Critically, the per-step decode time is typically much smaller than the TPOT SLO—as noted earlier, a 128K-token request incurs roughly 40.3ms per step, well below a typical 50ms SLO budget. This leaves a slack of tens of milliseconds after each decode step, which can be exploited to schedule additional computation without violating the SLO.

The above analysis highlights the inefficiencies caused by imbalanced, long-tailed request distributions and motivates us to design an optimized scheduling method that improves both SLO attainment and system throughput.

\section{Design}

The design of \sys is designed to mitigate the inefficiencies caused by long-tail request distributions in disaggregated LLM serving systems, by introducing specialized scheduling mechanisms for each phase.

Figure~\ref{fig:arch} provides an architectural overview of \sys, illustrating how these two components work within a disaggregated serving framework.
On the prefill side (\S\ref{subsec:urgency-based-priority-scheduling}), \sys employs urgency-based priority scheduling. By predicting the prefill time for all queued requests and computing their urgency based on remaining SLO budget, the prefill scheduler greedily selects the most urgent requests to prefill, preventing long requests from starving short ones and maximizing overall TTFT SLO attainment.
On the decode side (\S\ref{subsec:slacked-guided-adaptive-scheduling}), \sys introduces slack-guided adaptive scheduling. By profiling per-step decode times and exploiting the slack between step time and the TPOT SLO, the decode scheduler greedily packs short requests to maximize throughput while ensuring that every request meets its token delivery deadline.
We detail each component in the following subsections.

\begin{algorithm}[tb]
  \caption{Urgency-Based Prefill Scheduling}
  \label{algo:prefill}
  \begin{algorithmic}[1]
    \STATE {\bfseries Input:} request queue $\mathcal{Q}$, fixed chunk-size $C$, current time $t_{\text{now}}$
    % \STATE {\bfseries Output:} selected request batch $\mathcal{B}$
    \STATE $\mathcal{B} \leftarrow \emptyset$, \; $c_{\text{used}} \leftarrow 0$
    \STATE $\triangleright$ \textit{Compute normalized urgency for each request}
    \STATE Sort $\mathcal{Q}$ by arrival time $r.t_{\text{arrive}}$ (FCFS order)
    \FOR{each request $r \in \mathcal{Q}$}
      \STATE $\hat{t}_r^{\text{finish}} \leftarrow \textsc{PredictPrefillFinishTime}(\mathcal{Q}, r, t_{\text{now}})$
      \STATE $t_r^{\text{slack}} \leftarrow r.\text{SLO}_{\text{TTFT}} - (\hat{t}_r^{\text{finish}} - r.t_{\text{arrive}})$
      \STATE $u_r^{normalized} \leftarrow \frac{t_r^{\text{slack}}}{r.\text{SLO}_{\text{TTFT}}} \,/\, r.\text{input\_len}$ \hfill $\triangleright$ normalized urgency score
    \ENDFOR
    \STATE $\triangleright$ \textit{Select the most urgent requests to prefill}
    \STATE Sort $\mathcal{Q}$ in descending order of $u_r^{normalized}$
    \FOR{each request $r$ in sorted $\mathcal{Q}$}
      \IF{$c_{\text{used}} + r.\text{input\_len} \leq C$}
        \STATE $\mathcal{B} \leftarrow \mathcal{B} \cup \{r\}$
        \STATE $c_{\text{used}} \leftarrow c_{\text{used}} + r.\text{input\_len}$
      \ELSE
        \STATE $\mathcal{B} \leftarrow \mathcal{B} \cup \{r[:C-c_{\text{used}}]\}$
        \STATE $C \leftarrow c_{\text{used}}$
      \ENDIF
    \ENDFOR
    \STATE $t_{\text{actual}} \leftarrow$ \textsc{DecodeStep}($\mathcal{B}$)
    \STATE \textsc{UpdateThroughput}($\sum_{r\in\mathcal{B}}r.seq\_len$, $t_{\text{actual}}$)
    % \STATE \textbf{return} $\mathcal{B}$
  \end{algorithmic}
\end{algorithm}

\subsection{Urgency-based Priority Scheduling}
\label{subsec:urgency-based-priority-scheduling}

\textbf{Limitations of existing approaches.} Existing disaggregated serving systems typically employ first-come-first-served (FCFS) scheduling on the prefill side~\cite{zheng2024sglang, kwon2023efficient}. While simple, FCFS is highly vulnerable to head-of-line (HOL) blocking: a long request that arrives first can monopolize the prefill stage and delay all subsequent short requests. As analyzed in \S\ref{subsec:request-imbalance}, when long and short requests arrive in sequence, the TTFT SLO begins to be violated even at low request rate. One natural alternative is shortest-job-first (SJF) scheduling, which prioritizes short requests and can substantially improve TTFT attainment. However, SJF is impractical in online serving scenarios: long requests may be indefinitely starved, as newly arriving short requests continuously jump ahead in the queue, causing unbounded delays for long requests and eventual SLO violations for them.

\begin{align}
urgency_R = \frac{SLO_{TTFT} - (T^R_{finish} - T^R_{arrival})}{SLO_{TTFT}}
\label{eq:urgency}
\end{align}

\begin{algorithm}[tb]
  \caption{\textsc{PredictPrefillFinishTime}: FCFS-based prefill finish time estimation}
  \label{algo:predict-prefill}
  \begin{algorithmic}[1]
    \STATE {\bfseries Input:} queue $\mathcal{Q}$ (sorted by $t_{\text{arrive}}$), target request $r$, current time $t_{\text{now}}$
    \STATE {\bfseries Output:} estimated prefill finish time $\hat{t}_r^{\text{finish}}$
    \STATE $t_{\text{cursor}} \leftarrow t_{\text{now}}$ \hfill $\triangleright$ simulated clock
    \FOR{each request $r' \in \mathcal{Q}$ with $r'.t_{\text{arrive}} \leq r.t_{\text{arrive}}$}
      \STATE $n_{r'} \leftarrow r'.\text{remaining\_prefill\_tokens}$
      \STATE $\hat{d}_{r'} \leftarrow n_{r'} \,/\, \mu_{\text{prefill}}$ \hfill $\triangleright$ estimated prefill duration via profiled throughput $\mu_{\text{prefill}}$
      \STATE $t_{\text{cursor}} \leftarrow \max(t_{\text{cursor}},\; r'.t_{\text{arrive}}) + \hat{d}_{r'}$
    \ENDFOR
    \STATE \textbf{return} $t_{\text{cursor}}$
  \end{algorithmic}
\end{algorithm}

\textbf{Our approach.} To address this challenge, we propose urgency-based priority scheduling, whose pseudocode is presented in Algorithm~\ref{algo:prefill}. \sys adopts chunked prefill~\cite{agrawal2023sarathi} to bound the peak memory footprint when processing extremely long sequences, splitting them into fixed-size chunks. Unlike existing FCFS-based approaches, \sys selects certain chunks at the beginning of each prefill step. Specifically, \sys computes an urgency score (Line 7-8) for each queued request before every prefill scheduling decision. The urgency score is defined by Equation~\ref{eq:urgency}, where the numerator captures the ratio of the slack between prefill finish time to the TTFT SLO under FCFS scheduling policy. If this ratio is greater than 0, there is slack that can be exploited without violating the SLO; otherwise, the request has already exceeded its TTFT SLO under FCFS order. \sys normalizes the urgency score by request length (Line 8), such that shorter requests naturally receive higher priority than longer ones at comparable finish times. The scheduler then greedily selects the chunks of the requests with the highest normalized urgency for prefill (Line 11-20), where the total length of these chunk is less than the fixed chunk-size.

Additionally, as shown in Algorithm~\ref{algo:predict-prefill}, \sys re-estimates the finish time of each request at the beginning of each step to avoid the impact of the prefix cache hit ratio, which has a significant impact on the number of tokens to prefill. \sys maintains a running estimate of the average prefill throughput (tokens per second, TPS) on the prefill instance (Line 6). This allows the system to accurately predict the execution time of a queued request by dividing its token count by the current throughput estimate, and this prediction is used in computing the urgency (Line 6 of Algorithm~\ref{algo:prefill}).

This design directly addresses the HOL blocking problem. Consider a scenario where a 128K-token request arrives first, followed shortly by a few 8K-token request, as shown in Figure~\ref{fig:arch}. Under FCFS, the short request must wait for the entire long prefill to complete, likely violating its TTFT SLO. In contrast, \sys computes the urgency scores for both requests. Even though the long request arrived earlier and has waited longer, the short request's urgency score is amplified by its shorter length, allowing it to be selected first. As a result, the short request meets its SLO, while the long request retains a reasonable urgency score since it still has slack before its own TTFT SLO expires. This mechanism naturally balances SLO attainment across requests of varying lengths without risking starvation.

\subsection{Slacked-guided Adaptive Scheduling}
\label{subsec:slacked-guided-adaptive-scheduling}

\begin{algorithm}[tb]
	\caption{Slack-guided Adaptive Decode Scheduling}
	\label{algo:decode}
	\begin{algorithmic}[1]
		\STATE {\bfseries Input:} active request set $\mathcal{D}$, TPOT SLO $T_{\text{SLO}}$, lookup table $\textsc{LUT}[\textit{bsz}, \textit{seq\_len}]$
		% \STATE {\bfseries Output:} requests to decode this step $\mathcal{B}$
		\STATE $\triangleright$ \textit{Compute slack for each active request}
		\FOR{each request $r \in \mathcal{D}$}
		\STATE $t_r^{\text{elapsed}} \leftarrow t_{\text{now}} - r.t_{\text{first\_token}}$
		\STATE $s_r \leftarrow T_{\text{SLO}} \cdot (r.n_{\text{gen}} + 1) - t_r^{\text{elapsed}} - LUT[1, r.seq\_len]$ \hfill $\triangleright$ slack before SLO violation
		\ENDFOR
		\STATE $s_{\min} \leftarrow \min_{r \in \mathcal{D}} \{s_r\}$
		\STATE $\triangleright$ \textit{Partition requests by slack and throughput gain}
		\STATE $\mathcal{B} \leftarrow \emptyset$, \; $\mathcal{R}_{\text{delay}} \leftarrow \emptyset$, \; $t_{\text{cur}} \leftarrow 0$
		\STATE Sort $\mathcal{D}$ in ascending order of r.seq\_len
		\FOR{each request $r \in \text{sorted}\ \mathcal{D}$}
		\STATE $t_r^{\text{step}} \leftarrow \textsc{LUT}[|\mathcal{B} \cup \{r\}|, \; r.\text{seq\_len}]$
		\IF{$t_r^{\text{step}} \leq s_{\min}$ \textbf{and} $(|\mathcal{B}| = 0$ \textbf{or} $\frac{|\mathcal{B}|+1}{t_r^{\text{step}}} > \frac{|\mathcal{B}|}{t_{\text{cur}}})$}
		\STATE $\mathcal{B} \leftarrow \mathcal{B} \cup \{r\}$, \; $t_{\text{cur}} \leftarrow t_r^{\text{step}}$
		\ELSE
		\STATE $\mathcal{R}_{\text{delay}} \leftarrow \mathcal{R}_{\text{delay}} \cup \{r\}$
		\ENDIF
		\ENDFOR
		\IF{$\mathcal{B} = \emptyset$}
		\STATE $\mathcal{B} \leftarrow \mathcal{D}$ \hfill $\triangleright$ no slack to exploit; decode all
		\ENDIF
		\STATE $\triangleright$ \textit{Execute decode step and update LUT}
		\STATE $t_{\text{actual}} \leftarrow$ \textsc{DecodeStep}($\mathcal{B}$)
		\STATE \textsc{UpdateLUT}($|\mathcal{B}|$, $\max_{r \in \mathcal{B}}\{r.\text{seq\_len}\}$, $t_{\text{actual}}$)
		\FOR{each $r \in \mathcal{B}$}
		\STATE $r.n_{\text{gen}} \leftarrow r.n_{\text{gen}} + 1$
		\IF{$r$ generated EOS token}
		\STATE Remove $r$ from $\mathcal{D}$; return result to client
		\ENDIF
		\ENDFOR
		% \STATE \textbf{return} $\mathcal{B}$
	\end{algorithmic}
\end{algorithm}

\textbf{Limitations of existing approaches.} Existing disaggregated serving systems employ continuous batching on the decode side, which eagerly adds new requests to the current batch as soon as slots become available~\cite{kwon2023efficient}. However, this approach ignores the straggler effect: when requests of vastly different lengths coexist in a batch, all requests are paced by the slowest one. As quantified in \S\ref{subsec:request-imbalance}, the per-step decode time for a 128K-token request is roughly 4$\times$ that of a 8K-token request. Even if requests are initially grouped by input length, their output lengths can diverge significantly during generation—a short-prompt request may produce a very long output and vice versa—making static grouping insufficient to avoid stragglers over the course of decoding.

\begin{align}
    slack_R = &(N+1)\times SLO_{TPOT} - (T_{cur}-T^{finish}_{prefill}) -\notag
            \\&LUT(SeqLen_R, BatchSize) \label{eq:slack}
\end{align}

\textbf{Our approach.} To address this issue, we propose slack-guided adaptive batching, as is shown in Algorithm~\ref{algo:decode}. \sys constructs a lookup table (LUT) mapping batch size and sequence length to per-step decode time (Line 1). This LUT is profiled offline once per model-hardware configuration and incurs no runtime overhead. Before each decode step, \sys computes a slack value for every active request, defined as the remaining time budget before the next token must be delivered to meet the TPOT SLO. As formulated in Equation~\ref{eq:slack}, the slack is calculated based on the request's prefill finish time, current sequence length, number of tokens generated so far ($N$), and the TPOT SLO. \sys identifies the minimum slack among all active requests and compares it against the per-step decode times obtained from the LUT (Line 3-7). If there exist requests whose decode step time is smaller than this minimum slack, and packing them can improve the decode throughput, \sys selectively executes only those requests and delays the remaining ones for this step (Line 11-18); otherwise, it proceeds to decode all requests together (Line 19-21).

The profiled LUT captures the mean decode step time over 100 profiling runs per configuration. However, step times could fluctuate at runtime due to hardware variability, power throttling, or interference from other processes. To remain robust to these dynamics, \sys continuously updates the LUT at runtime (Line 23-24) based on observed decode step times. Specifically, after each decode step, \sys updates the corresponding LUT entry using the historical mean of all observed step times for that batch size and sequence length combination, ensuring the scheduling decisions are accurate and reliable.

This design eliminates unnecessary waiting caused by stragglers at each step. For instance, as illustrated in the right part of Figure~\ref{fig:arch}, suppose the batch contains a few short requests with short decode time and a long request with long decode time, under a 50ms TPOT SLO. When the slack is enough (Step 3), \sys recognizes that it can execute the short request within the SLO budget before the long request must be served. Rather than forcing the short request to idle for each step alongside the straggler, \sys decouples their execution, allowing the short request to progress rapidly and buffering its output tokens, thereby significantly improving decode throughput without violating any SLO.
\section{Experiment}

\subsection{Experiment Settings}
\label{subsec:experiment-settings}

\textbf{Model.} We evaluate \sys on Minimax-M2.5 (M2.5)~\cite{minimax-m2.5} and Qwen3-VL-8B (Qwen)~\cite{bai2025qwen3}. M2.5 is a 229B-parameter text-only model served in FP8 precision, optimized for agentic coding and reasoning. This model represents the frontier of large-scale Mixture-of-Experts architectures and achieves state-of-the-art results across a wide range of benchmarks. Qwen is a compact yet powerful vision-language model from the Qwen series, built upon an 8-billion-parameter dense architecture. It integrates a visual encoder with a large language model via a cross-attention mechanism, enabling strong multimodal understanding and reasoning capabilities.

\textbf{Dataset.} We use a production online serving trace collected from a real-world LLM service, containing 1000 requests with a pronounced long-tail distribution in sequence lengths. This dataset allows us to evaluate \sys under realistic production workloads.

\textbf{Testbed.} All experiments are conducted on a single server equipped with 8× NVIDIA H200 SXM GPUs, 192 Intel Xeon Platinum 8558 CPU cores, and 2 TB of host memory. The NVLink bandwidth is 900 GB/s~\cite{h200}. We deploy the disaggregated architecture with 4 GPUs dedicated to the prefill phase and 4 GPUs dedicated to decode, both configured with tensor parallelism degree 4 (TP4) for M2.5; one GPU dedicated to prefill and one GPU dedicated to decode for Qwen. The server uses NVLink and NVSwitch for intra-node GPU communication.

\textbf{Baselines.} We compare \sys against DistServe~\cite{zhong2024distserve}, which applies the FCFS strategy and employs the same 4P+4D or 1P+1D disaggregated setup with pure TP for each phase, serving as a direct architectural baseline to isolate the benefits of our scheduling policies. DistServe uses default FCFS prefill scheduling and continuous batching for decode.

\textbf{Metrics.} We evaluate \sys along four dimensions: (1) \textit{TTFT SLO attainment} captures the fraction of requests whose time-to-first-token meets the prefill-side latency target; (2) \textit{TPOT SLO attainment} measures the fraction of requests where the mean inter-token latency falls within the SLO budget; (3) \textbf{End-to-end SLO attainment} measures the fraction of requests that meets both TTFT and TPOT SLOs; (4) \textit{Decode throughput} quantifies the per-request decode speed in tokens per second, reflecting decode-side efficiency.

\textbf{SLO Configuration.} We set the TTFT SLO to 8s and the TPOT SLO to 50ms for M2.5, and set the TTFT SLO to 4s and the TPOT SLO to 15ms for Qwen. These values are chosen to reflect typical production requirements: the 8s TTFT SLO corresponds to a commonly acceptable initial latency for interactive applications, while the 50ms TPOT SLO maps to a human-perceptible token generation rate of 20 tokens per second.

\subsection{End-to-End SLO Attainment}

%\begin{figure*}[ht]
%	\centering					
%	\includegraphics[width=0.95\textwidth]{fig/test-SLO-attainment.pdf}
%	\caption{End-to-end SLO attainment of each model and dataset under differnet QPS.}
%	\label{fig:e2e-slo-attainment}
%\end{figure*}

%\begin{figure}[t!]
%	\centering					
%	\includegraphics[width=0.4\textwidth]{fig/glmdata-slo-e2e.pdf}
%	\caption{End-to-end SLO attainment of Minimax M2.5 and online dataset under differnet QPS.}
%	\label{fig:e2e-slo-attainment}
%\end{figure}

\begin{figure}[t!]
	\centering
	\begin{subfigure}[b]{0.24\textwidth}
		\centering
		\includegraphics[width=\textwidth]{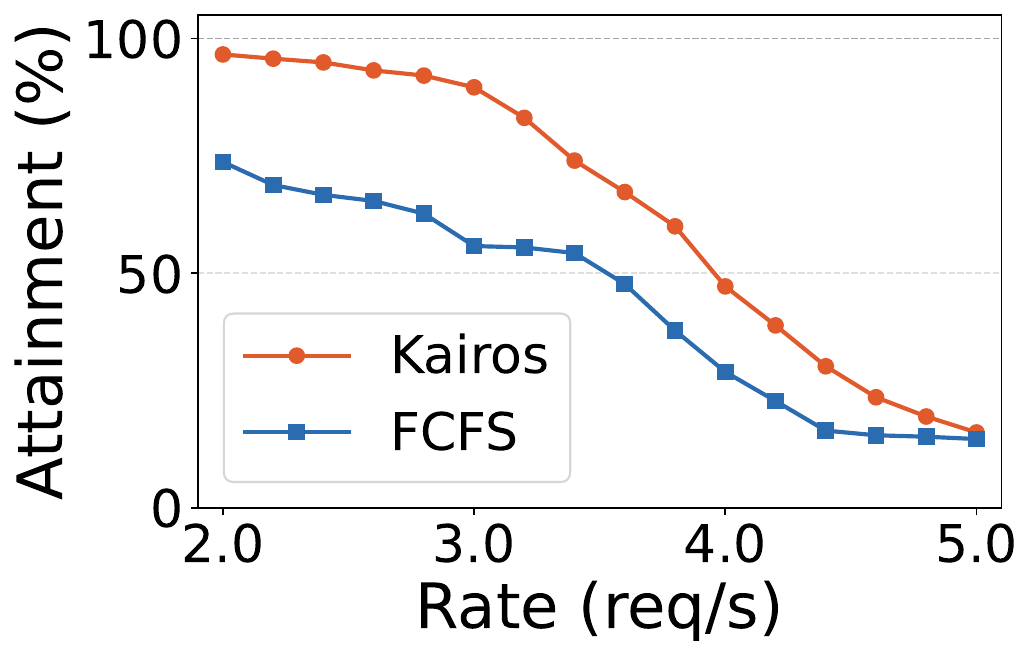}
		\caption{M2.5 end-to-end SLO attainment.}
		\label{fig:e2e-slo-attainment-m25}
	\end{subfigure}%
	\hfill
	\begin{subfigure}[b]{0.24\textwidth}
		\centering
		\includegraphics[width=\textwidth]{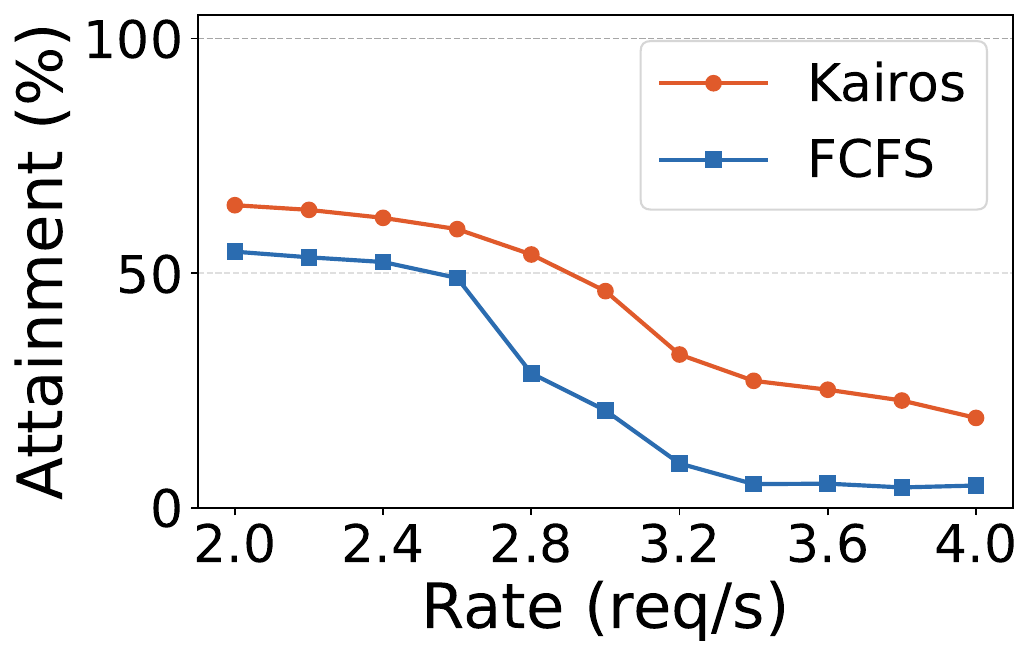}
		\caption{Qwen end-to-end SLO attainment.}
		\label{fig:e2e-slo-attainment-qwen}
	\end{subfigure}
	\caption{End-to-end SLO attainment under different QPS.}
	\label{fig:e2e-slo-attainment}
\end{figure}

Figure~\ref{fig:e2e-slo-attainment-m25} presents the combined SLO attainment (both TTFT and TPOT satisfied per request) as a function of request rate for M2.5 on the online dataset. \sys consistently outperforms DistServe across all QPS levels. At low load (QPS$\leq$2.8), \sys achieves 92--97\% combined attainment versus 63--74\% for DistServe, a gap of 22\%--29\%. At the critical inflection point of QPS$=$3.0, \sys reaches 89.6\% combined attainment while DistServe collapses to 55.8\%, an improvement of 33.8\%. As QPS increases beyond 3.0, both systems degrade, but \sys sustains consistently higher attainment: at QPS$=$4.0, \sys achieves 47.2\% versus 29.0\% for DistServe (+18.2\%), and at QPS$=$4.2, 38.9\% versus 22.8\% (+16.1\%).

The performance gap is especially striking on the TTFT dimension: DistServe TTFT attainment drops sharply from 100\% to 76.1\% at QPS$=$3.0, while \sys maintains 100\% TTFT attainment through QPS$=$3.0 and does not fall below 90\% until QPS$=$3.6. This demonstrates that the urgency-based priority scheduling effectively prevents head-of-line blocking, keeping short requests from being stalled behind long ones under the FCFS discipline.

Figure~\ref{fig:e2e-slo-attainment-qwen} reports the same combined SLO attainment for Qwen under the tighter 4s TTFT / 15ms TPOT budget. \sys again dominates DistServe throughout the entire QPS sweep. At light load (QPS$=$2.0--2.6), \sys attains 59.4\%--64.5\% end-to-end SLO versus 49.0\%--54.6\% for DistServe, an absolute gain of 9.4--10.9 percentage points. At the saturation knee of QPS$=$3.0, DistServe collapses to 20.8\% while \sys sustains 46.2\%, delivering the largest gap of \textbf{+25.4 pp}. The advantage persists deep into the overloaded regime: at QPS$=$3.2 the gap is +23.2 pp (9.5\% vs.\ 32.7\%), and even at QPS$=$4.0 \sys still delivers 19.2\% versus 4.8\% for DistServe (+14.4 pp, a $4.0\times$ relative improvement). These results confirm that \sys's benefits generalize from the large-scale MoE regime of M2.5 to the compact multimodal setting of Qwen.

\subsection{TTFT SLO Attainment}

%\begin{figure*}[ht]
%	\centering					
%	\includegraphics[width=0.95\textwidth]{fig/test-SLO-attainment.pdf}
%	\caption{TTFT SLO attainment of each model and dataset under differnet QPS.}
%	\label{fig:ttft-slo-attainment}
%\end{figure*}

%\begin{figure}[t!]
%	\centering					
%	\includegraphics[width=0.4\textwidth]{fig/glmdata-slo-ttft.pdf}
%	\caption{TTFT SLO attainment of Minimax M2.5 and online dataset under differnet QPS.}
%	\label{fig:ttft-slo-attainment}
%\end{figure}

\begin{figure}[t!]
	\centering
	\begin{subfigure}[b]{0.24\textwidth}
		\centering
		\includegraphics[width=\textwidth]{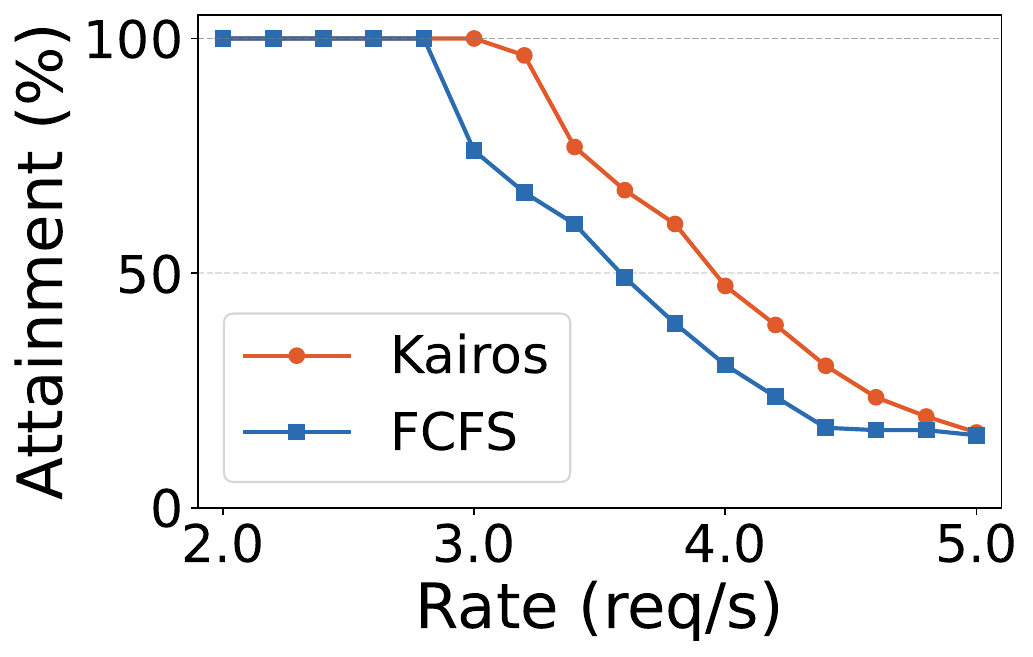}
		\caption{M2.5 TTFT SLO attainment.}
		\label{fig:ttft-slo-attainment-m25}
	\end{subfigure}%
	\hfill
	\begin{subfigure}[b]{0.24\textwidth}
		\centering
		\includegraphics[width=\textwidth]{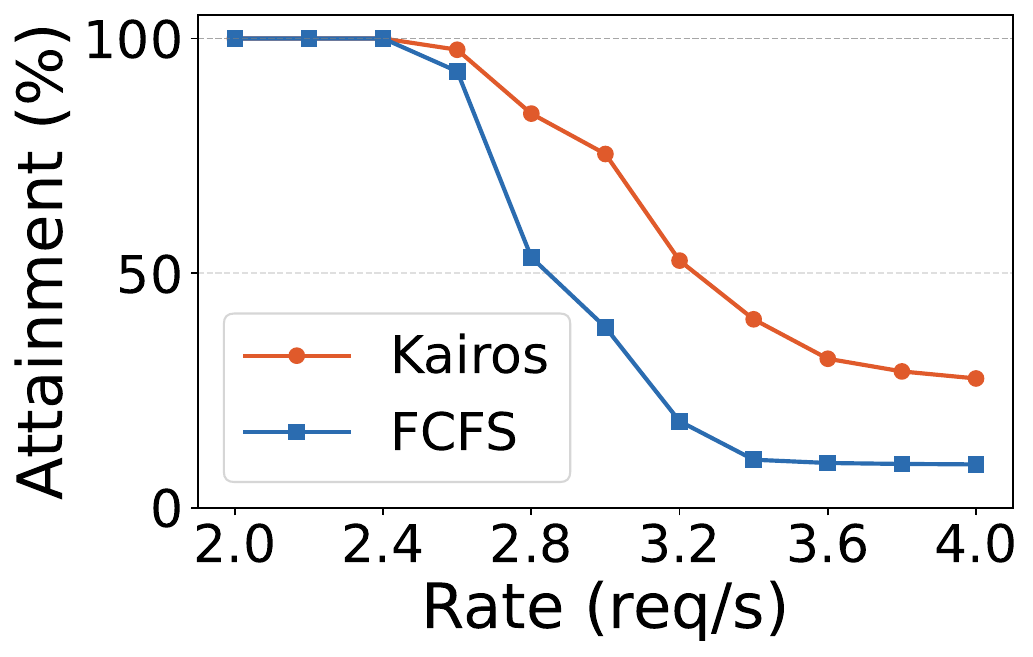}
		\caption{Qwen TTFT SLO attainment.}
		\label{fig:ttft-slo-attainment-qwen}
	\end{subfigure}
	\caption{TTFT SLO attainment under different QPS.}
	\label{fig:ttft-slo-attainment}
\end{figure}

Figure~\ref{fig:ttft-slo-attainment-m25} shows the TTFT SLO attainment as QPS increases for M2.5. At low QPS ($\leq$2.8), both systems achieve 100\% TTFT attainment, as the system is under-loaded and all requests are served promptly. As QPS climbs, DistServe degrades sharply: its TTFT attainment drops to 76.1\% at QPS$=$3.0, 60.5\% at QPS$=$3.4, and 15.5\% at QPS$=$5.0. In contrast, \sys maintains 100\% TTFT attainment at QPS$=$3.0 and degrades more gracefully, reaching 76.9\% at QPS$=$3.4 and 16.1\% at QPS$=$5.0. At QPS$=$3.0, \sys improves TTFT attainment by 23.9\% over DistServe.

The superior TTFT performance of \sys stems from its urgency-based priority scheduling on the prefill side. When a long request arrives ahead of short ones, \sys recognizes that the short requests are more urgent relative to their SLO deadline and schedules them first, preventing their TTFT from violating the SLO. DistServe, by contrast, enforces strict arrival order, so short requests are helplessly blocked behind long ones, causing TTFT violations to accumulate as QPS increases.

Figure~\ref{fig:ttft-slo-attainment-qwen} shows the TTFT results for Qwen. Both systems achieve 100\% attainment up to QPS$=$2.4, but their trajectories diverge sharply beyond this point. At QPS$=$2.6, DistServe already drops to 93.0\% while \sys still holds 97.6\%. The gap widens dramatically at QPS$=$2.8, where DistServe plunges to 53.4\% while \sys retains 84.0\% (+30.6 pp). The maximum improvement occurs at QPS$=$3.0: DistServe reaches only 38.5\%, whereas \sys sustains 75.4\%, an absolute gain of \textbf{+36.9 pp} and a nearly $2\times$ relative improvement. Even at QPS$=$3.2--4.0, where DistServe collapses below 20\%, \sys maintains 27.6\%--52.7\% TTFT attainment, corresponding to +18.3 to +34.2 pp of improvement. This confirms that urgency-based prefill scheduling is equally effective at protecting short vision-language requests from head-of-line blocking under a tight 4s TTFT budget.

\subsection{TPOT SLO Attainment}

%\begin{figure*}[ht]
%	\centering					
%	\includegraphics[width=0.95\textwidth]{fig/test-SLO-attainment.pdf}
%	\caption{TPOT SLO attainment of each model and dataset under differnet QPS.}
%	\label{fig:tpot-slo-attainment}
%%	\vspace{-10pt}
%\end{figure*}

%\begin{figure}[t]
%	\centering					
%	\includegraphics[width=0.4\textwidth]{fig/glmdata-slo-tpot.pdf}
%	\caption{TTFT SLO attainment of Minimax M2.5 and online  under differnet QPS.}
%	\label{fig:tpot-slo-attainment}
%\end{figure}

\begin{figure}[t!]
	\centering
	\begin{subfigure}[b]{0.24\textwidth}
		\centering
		\includegraphics[width=\textwidth]{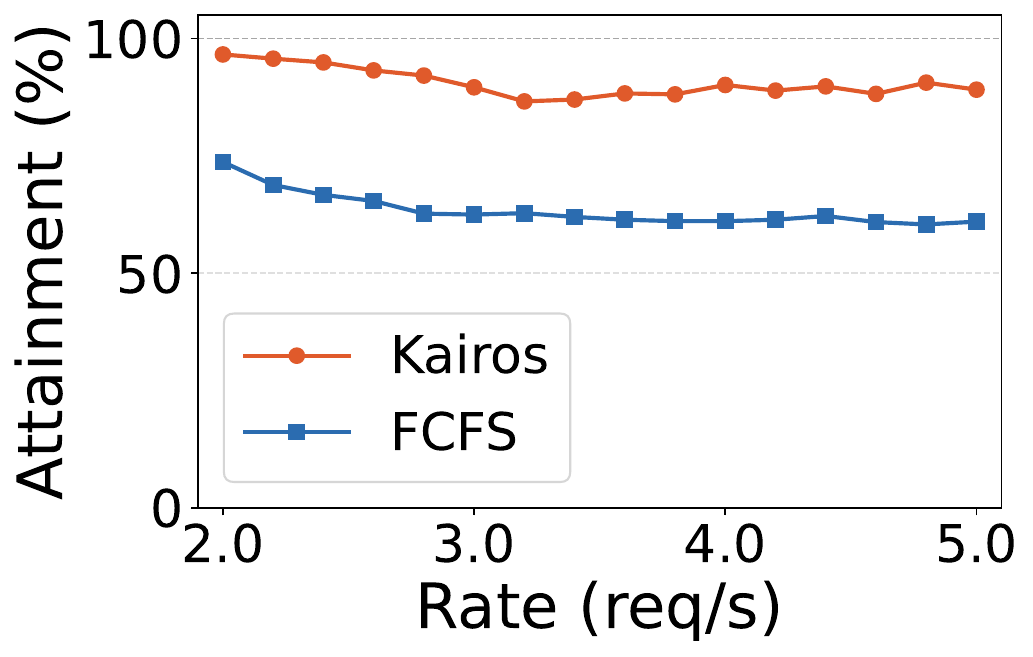}
		\caption{M2.5 TPOT SLO attainment.}
		\label{fig:tpot-slo-attainment-m25}
	\end{subfigure}%
	\hfill
	\begin{subfigure}[b]{0.24\textwidth}
		\centering
		\includegraphics[width=\textwidth]{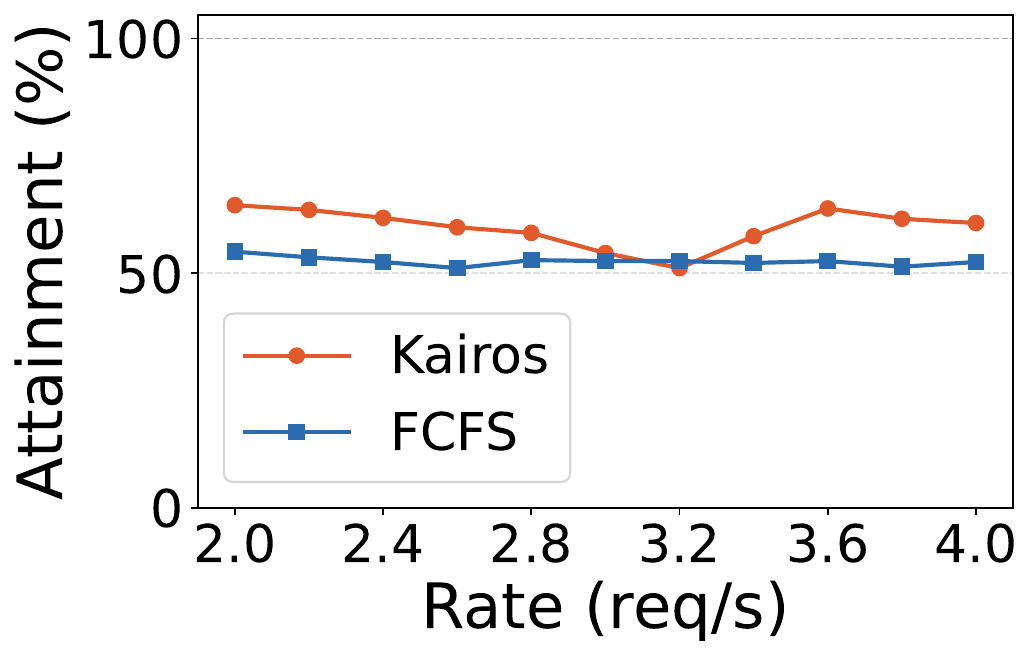}
		\caption{Qwen TPOT SLO attainment.}
		\label{fig:tpot-slo-attainment-qwen}
	\end{subfigure}
	\caption{TPOT SLO attainment under different QPS.}
	\label{fig:tpot-slo-attainment}
\end{figure}

Figure~\ref{fig:tpot-slo-attainment-m25} presents the TPOT SLO attainment as QPS increases for M2.5. \sys substantially outperforms DistServe across the entire QPS range. While the TPOT attainment of DistServe stagnates between 60\% and 74\% regardless of load---indicating a fundamental inability to meet the 50ms TPOT target for a large fraction of requests---\sys achieves 92--97\% at low QPS (2.0--2.8) and remains above 83\% even at QPS$=$3.2. At QPS$=$2.0, \sys improves TPOT attainment from 73.7\% to 96.6\%, a gain of 22.9\%. At QPS$=$3.0, the gap is 27.1\% (89.6\% vs. 62.5\%).

The structural disadvantage of DistServe on TPOT arises from its continuous batching policy, which batches all active decode requests together regardless of their individual TPOT slack. Long requests with many remaining tokens inflate the decode batch and slow down the per-step latency for all requests. \sys's slack-guided adaptive batching avoids this: it adjusts batch composition based on per-request TPOT slack, allowing short or time-critical requests to progress quickly without being stalled by long-tail stragglers. This enables \sys to keep the decode step time well within the 50ms budget for the vast majority of requests.

Figure~\ref{fig:tpot-slo-attainment-qwen} presents the TPOT results for Qwen under the stringent 15ms per-token budget. DistServe's TPOT attainment is essentially flat around 51--55\% regardless of load, again exhibiting a structural ceiling imposed by unmanaged decode batching. \sys consistently exceeds this ceiling, sustaining 51.1\%--64.5\% across the entire QPS range. The gains are pronounced across both low and high load: at QPS$=$2.0 \sys reaches 64.5\% versus 54.6\% (+9.9 pp), at QPS$=$2.2 the gap is +10.1 pp (53.4\% vs.\ 63.5\%), and the maximum gap of \textbf{+11.2 pp} occurs at QPS$=$3.6 (52.6\% vs.\ 63.8\%). Even under heavy overload (QPS$=$3.8--4.0), \sys retains a +8--10 pp advantage. Because the 15ms TPOT bound is inherently more difficult to meet than the 50ms M2.5 target, the absolute attainment levels are lower, but the structural advantage of slack-guided batching over vanilla continuous batching is preserved.

\subsection{Decode Throughput}

%\begin{figure*}[ht]
%	\centering					
%	\includegraphics[width=0.95\textwidth]{fig/test-SLO-attainment.pdf}
%	\caption{Decode throughput of each model and dataset under differnet QPS.}
%	\label{fig:decode-throughput}
%%	\vspace{-10pt}
%\end{figure*}

%\begin{figure}[t]
%	\centering					
%	\includegraphics[width=0.4\textwidth]{fig/glmdata-decode-throughput.pdf}
%	\caption{TTFT SLO attainment of Minimax M2.5 and online dataset under differnet QPS.}
%	\label{fig:decode-throughput}
%\end{figure}

\begin{figure}[t!]
	\centering
	\begin{subfigure}[b]{0.24\textwidth}
		\centering
		\includegraphics[width=\textwidth]{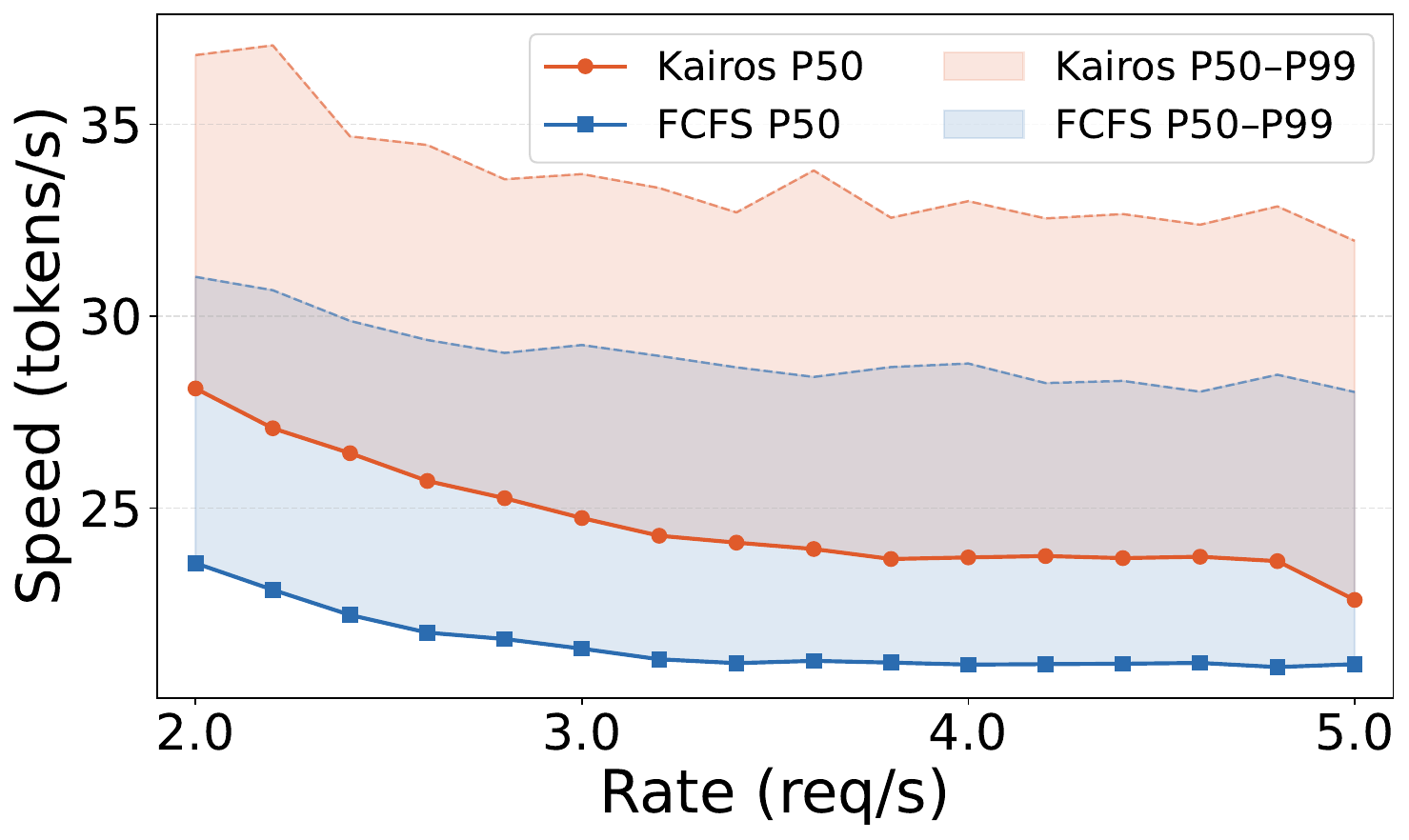}
		\caption{M2.5 decode-throughput.}
		\label{fig:decode-throughput-m25}
	\end{subfigure}%
	\hfill
	\begin{subfigure}[b]{0.24\textwidth}
		\centering
		\includegraphics[width=\textwidth]{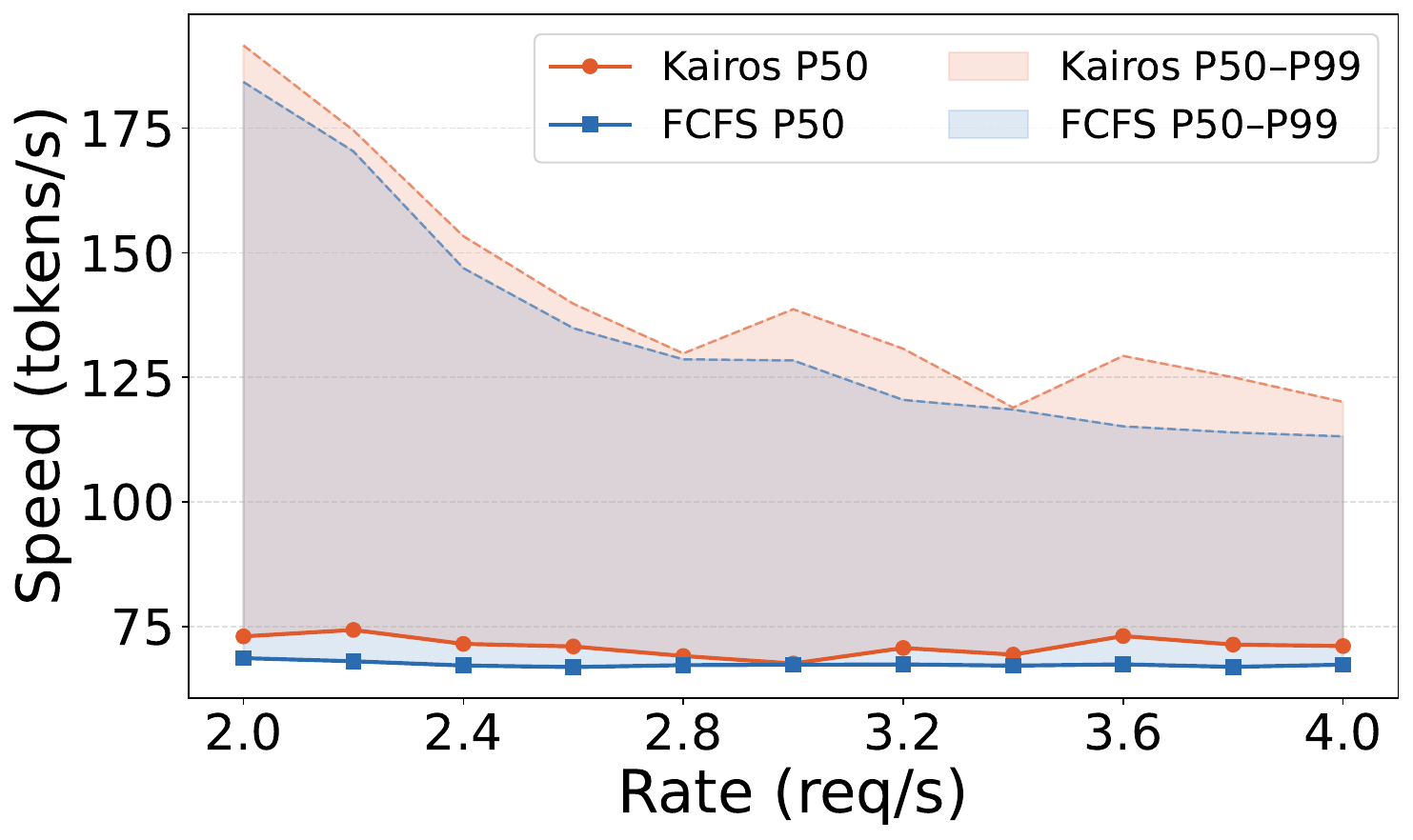}
		\caption{Qwen decode-throughput.}
		\label{fig:decode-throughput-qwen}
	\end{subfigure}
	\caption{Decode throughput under different QPS.}
	\label{fig:decode-throughput}
\end{figure}

Figure~\ref{fig:decode-throughput-m25} presents the per-request decode throughput (tokens per second) as a function of QPS for M2.5. \sys achieves consistently higher per-request decode throughput than DistServe across all QPS levels. At QPS$=$2.0, the median decode throughput is 28.1 tok/s for \sys versus 23.6 tok/s for DistServe, a relative improvement of \textbf{19.3\%}. Across QPS$=$2.0--4.8, \sys delivers a median decode throughput improvement of \textbf{13--19\%} over DistServe. Even at QPS$=$5.0, \sys maintains a P50 of 22.6 tok/s versus 20.9 tok/s for DistServe (+8.0\%).

Since prefill is compute-bound and our urgency-based scheduling only reorders request execution without altering the total computation, prefill throughput remains unchanged compared to the baseline. The decode throughput gains therefore reflect the direct benefit of slack-guided adaptive batching: by prioritizing requests with tight TPOT budgets and packing decode batches more efficiently, \sys keeps the decode GPUs highly utilized while serving more tokens per second to each individual request. These gains are most pronounced at low-to-moderate QPS, where the slack-guided policy has the most room to differentiate request priorities before the system becomes saturated.

We note that the absolute decode throughput values (20--28 tokens/s) are lower than the peak decode throughput achievable on this hardware in isolation. This is primarily due to GPU memory constraints on the decode node. The decode instance must maintain KV caches for all concurrently active requests; at high concurrency, the KV cache memory is exhausted and new requests cannot be admitted, which caps the maximum batch size and, consequently, the aggregate decode throughput. As a result, the system operates at a moderate concurrency ceiling dictated by memory capacity rather than compute throughput, and the reported decode throughput reflects this memory-bound regime rather than the compute-bound peak.

Figure~\ref{fig:decode-throughput-qwen} reports per-request decode throughput for Qwen. \sys consistently outperforms DistServe at every QPS level on both the P50 and P99 tails. At QPS$=$2.0, \sys achieves a median decode throughput of 73.1 tok/s versus 68.7 tok/s for DistServe (+\textbf{6.4\%}), with the P99 tail improving from 184.3 to 191.6 tok/s. The relative gains hold as load increases: at QPS$=$3.6, \sys delivers a P50 of 73.1 tok/s versus 67.4 tok/s (+8.5\%) and a P99 of 129.3 versus 115.2 tok/s (+\textbf{12.2\%}). Across the full QPS$=$2.0--4.0 sweep, \sys yields a median throughput improvement of \textbf{0.3--9.3\%} on P50 and up to $\sim$12\% on P99. Because the Qwen deployment uses a single decode GPU rather than the 4-way TP configuration used for M2.5, the per-request tok/s values are natively higher, and the absolute throughput remains well within the compute-bound region even at high concurrency. This shows that slack-guided adaptive batching translates into decode-throughput gains across model scales and hardware configurations, complementing the SLO-attainment advantages reported above.

\section{Related Work}

\textbf{LLM serving systems.}
LLM inference consists of a prefill stage that processes the input prompt in parallel, and a decode stage that generates tokens autoregressively~\cite{achiam2023gpt, liu2024deepseek}. State-of-the-art serving systems include vLLM~\cite{kwon2023efficient}, which introduced PagedAttention and continuous batching; SGLang~\cite{zheng2024sglang}, which provides RadixAttention for prefix caching and efficient constrained decoding; and TensorRT-LLM~\cite{trtllm}, which leverages NVIDIA's compilation stack for optimized kernels. Disaggregated architectures, such as DistServe~\cite{zhong2024distserve} and Splitwise~\cite{patel2024splitwise}, separate prefill and decode across different GPU sets to reduce interference, while Sarathi~\cite{agrawal2023sarathi} introduces chunked prefill to bound peak memory for long sequences. Our work builds upon PD disaggregation and focuses on request scheduling, an orthogonal direction compatible with existing optimizations in these frameworks.

\textbf{Request scheduling.}
The limitations of FCFS scheduling have motivated extensive research. FastServe~\cite{wu2023fast} enables token-level preemption with a multi-level feedback queue. Preble~\cite{srivatsa2024preble} co-optimizes KV cache reuse and load balancing for prompt sharing. PARS~\cite{tao2025prompt} approximates SJF by learning pairwise rankings of output lengths. TRAIL~\cite{shahout2024don} predicts remaining generation length for constrained preemptive scheduling. AugServe~\cite{wang2025augserve} and MARS~\cite{shahout2024fast} incorporate runtime information for adaptive scheduling, while MemServe~\cite{hu2024memserve} manages distributed KV caches with global scheduling. K-LPM~\cite{dexter2025llm} addresses prefix reuse scheduling under TTFT constraints, and SageSched~\cite{gan2026sagesched} employs uncertainty-aware scheduling with lightweight output-length prediction.
In contrast, our work addresses request imbalance in disaggregated architectures without requiring output length prediction or expensive preemption. 
We exploit the predictability of prefill time and the slack between decode step time and SLO for lightweight, effective scheduling.

\textbf{Resource disaggregation.}
Resource disaggregation~\cite{guo2023mira, shan2018legoos} decouples hardware resources from monolithic infrastructure into independently managed pools, enabling flexible and efficient scaling. In LLM serving, disaggregation spans multiple dimensions: PD disaggregation (DistServe~\cite{zhong2024distserve}, Splitwise~\cite{patel2024splitwise}) separates prefill and decode to different instance; AF disaggregation (MegaScale-Infer~\cite{zhu2025megascale}) decouples attention from FFN layers; and EPD disaggregation (EPD~\cite{singh2024efficiently}, SpaceServe~\cite{li2025spaceserve}) separates encoders, prefill, and decode for multimodal models. Our work adopts PD disaggregation and is compatible with both AF and EPD approaches, allowing combined deployment for further gains.
\section{Conclusion}

We presented \sys, an SLO-aware scheduling system for disaggregated LLM serving that addresses the inefficiencies caused by long-tail request distributions. By leveraging two key observations—the predictability of prefill time and the slack between decode step time and TPOT SLO—\sys employs urgency-based priority scheduling on the prefill side and slack-guided adaptive batching on the decode side. Experiments show that \sys demonstrates significant improvements over state-of-the-art baselines in end-to-end SLO, TTFT and TPOT attainment, and decode throughput.

\bibliography{ref}
\bibliographystyle{icml2026}

\end{document}